\documentclass[epj,final]{svjour}
\usepackage{graphics,cite}
\newcommand{\PL}{Phys. Lett.\ }
\newcommand{\NP}{Nucl. Phys.\ }

\newcommand{\coll}{Collaboration}
\newcommand{\etal}{et al.}
 
\newcommand{\ff}        {\mbox{$F_{2}^{c\bar c}$}}

\begin{document}
\title{Open charm production in DIS at HERA}
\author{S.Chekanov
\thanks{For the H1 and ZEUS Collaborations} 
}
%
\institute{HEP Division, Argonne National Laboratory,
9700 S.Cass Avenue,
Argonne, \\ IL 60439 USA}
%
%
\abstract{
An overview of recent HERA results on   
inclusive production of 
$D^{*\pm}$ mesons in deep inelastic scattering is given. 
} 
\maketitle
\section{Introduction}
\label{intro}
The charm mass, $m_c$, is larger than the QCD dimensional scale $\Lambda_{QCD}$,
therefore, perturbative QCD is applicable at the scale  $m_c$.
In  deep inelastic scattering (DIS), 
another scale is the squared four-momentum transfer, $Q^2$,  
carried by the exchange photon.
The conventional QCD interpretation for $Q^2\sim m_c^2$ is that charm 
is determined solely by 
the gluon density, i.e. charm quarks are 
generated dynamically through the
boson-gluon fusion (BGF) process. 
Such approach is called the fixed-flavour-number scheme (FFNS).
For a sufficiently high $Q^2$,
this description may not be adequate, 
thus  the interplay between two independent scales,
$m_c$ and $Q$, embodies interesting QCD physics. 

Since the charm quarks are copiously produced via gluon splitting,
charm mesons can be used for testing different resummation techniques. 
The DGLAP resumption
is often  used together with the BGF process  
calculated up to next-to-leading-order
(NLO) QCD (the HVQDIS program \cite{pr:d57:2806}).
The BGF mechanism at leading-order QCD is implemented in    
parton-shower Monte Carlo (MC) models 
(AROMA \cite{cpc:101:1997:135} and  RAPGAP \cite{cpc:86:147}).   
Another description of the charm production
is based on the CCFM evolution \cite{ccfm}
as implemented in the CASCADE model \cite{epj:c19:351}.
This also starts from the BGF process but convoluted with the unintegrated
gluon density.

The ZEUS analysis discussed in this overview was 
performed with the data ($82$ pb$^{-1}$) taken from 1998 to
2000, when
electrons or positrons with energy $E_e =$ 27.6~GeV were
collided with  
protons of energy $E_p =$ 920~GeV. The  H1 Collaboration
uses 1996-1997 data,  when HERA operated with $E_p =$ 820~GeV. 
In this paper, only high-statistics results based on reconstructed $D^{*\pm}$ mesons 
are discussed
(other $D$ mesons are discussed in \cite{abs096}).   
The mesons were identified using the decay channel
$D^{*+}\to D^0\pi^{+}$ with the subsequent decay $D^0\to K^-\pi^+$
and corresponding antiparticle decay.
%
\begin{figure}
\begin{center}
\resizebox{0.33\textwidth}{!}{%
  \includegraphics{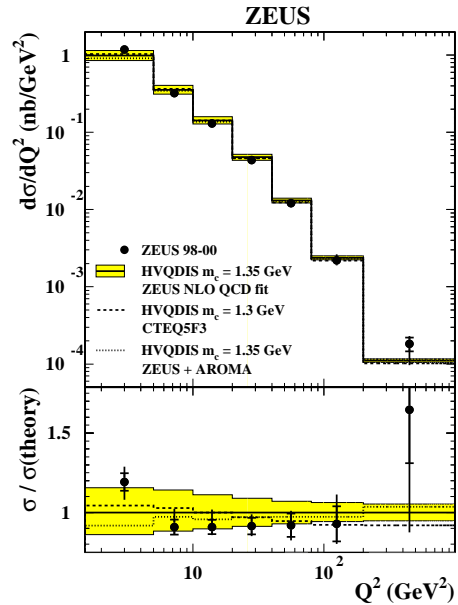} }
\caption{
Differential $D^{*\pm}$ cross section as a function of $Q^2$
for $0.02<y<0.7$, $1.5 < p_T(D^{*\pm}) < 15$~GeV and $|\eta(D^{*\pm})|<1.5$,
compared to the NLO QCD calculation of HVQDIS using the Peterson fragmentation
with $\epsilon=0.035$ and the hadronisation fraction $f(c\to D^{*})=0.235$ 
(solid line).
The renormalisation and factorisation scales were set to $\sqrt{Q^{2}/4 + m_{c}^{2}}$. 
Also shown are the NLO QCD predictions based on the CTEQ5F3 PDF (dashed-dotted line) and an
alternative hadronisation scheme (dotted line). 
The largest theoretical uncertainties (filled aria) are due to
the charm mass ($\pm 0.15$ GeV) and the scale 
variations (0.5$\mu_R$, 2$\mu_R$). 
}
\label{fig:1}       
\end{center}
\end{figure}

\section{Inclusive cross sections}
\label{sec:1}

Figure~1 shows the differential $D^{*\pm}$ cross section as a function
of $Q^2$ \cite{abs563}. The data falls by about four orders
of magnitude in the measured region. The NLO QCD calculations
based on the FFNS give a good agreement with the measurements 
up to $Q^2=1000$ GeV$^2$.
Predictions using an alternate PDF, CTEQ5F3, and
the Lund-string hadronisation extracted from AROMA, 
instead of the Peterson
fragmentation, are shown separately.

The pseudorapidity distribution of $D^{*\pm}$ 
mesons is  known to be particularly sensitive
to the underlying parton dynamics at small Bjorken $x$ \cite{baranov}.  
Figure~2 shows 
that the NLO calculation based on the
ZEUS NLO fit \cite{pr:d67:012007}, 
together with the Lund string fragmentation from AROMA, gives the best
description of the $\eta(D^{*\pm})$ cross section (and also better than the
prediction using  GRV98-HO, not shown). The NLO predictions with the CTEQ5F3, and with the
Peterson fragmentation, 
have distributions which are less shifted in the forward region.

\begin{figure}
\begin{center}
\resizebox{0.3\textwidth}{!}{%
  \includegraphics{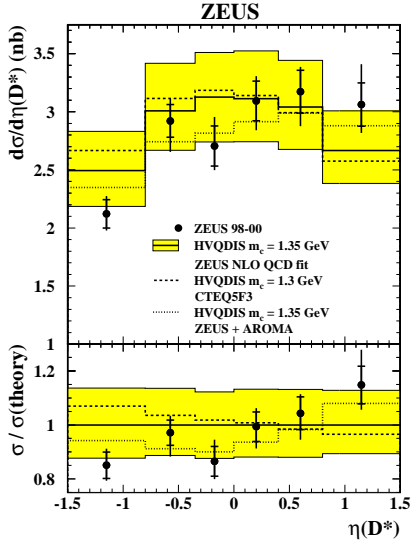} }
\caption{Differential $D^{*\pm}$ cross sections as a function of $\eta(D^{*\pm})$
compared to the NLO QCD calculation of HVQDIS. 
The cross section was measured for 
$Q^2>1.5$~GeV$^2$. 
Other details as for Fig.~1.}
\label{fig:2}       
\end{center}
\end{figure}

The H1 Collaboration also observes 
differences between data and NLO
QCD when the Peterson fragmentation and the 
CTEQ5F3 PDF is used for the NLO calculations \cite{abs074}, Fig.~3. 
The agreement with the data is better when the CASCADE model based on the
CCFM evolution is used. Since both HVQDIS and CASCADE use the Peterson fragmentation,
the difference between these models can be attributed to the parton dynamics
at low $x$.

In contrast, ZEUS uses the MC models with the 
Lund string fragmentation. In Fig.~4, AROMA and
CASCADE are compared  with the data. The CASCADE overestimates the data,
while AROMA is slightly below. 
The theoretical uncertainties,
which are expected to be larger than for the NLO calculations, 
were not estimated.    
At present, data are not
precise enough to distinguish  between the shapes of
the $\eta(D^{*\pm})$ distributions.

The resolved processes, 
in which the photon displays a hadronic
structure, can contribute to the $D^{*\pm}$ cross section at low $Q^2$. 
The resolved contribution, as implemented in the RAPGAP model, enhances
the $D^{*\pm}$ cross section in the rear direction  \cite{abs074}. Therefore,
it is unlikely that the observed 
differences between the NLO QCD and the data
can be explained by the resolved processes.

\begin{figure}
\begin{center}
\resizebox{0.3\textwidth}{!}{%
  \includegraphics{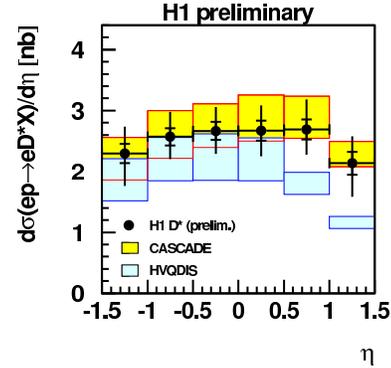}
}
\caption{Differential $D^{*\pm}$ cross section for $Q^2>2$ GeV$^2$ and $0.05<y<0.7$ 
as a function of $\eta(D^{*\pm})$ compared with the
CASCADE and HVQDIS programs. 
The kinematic range for the $p_T(D^{*\pm})$ as for Fig.1.  
Both theoretical 
predictions are based on the Peterson fragmentation with the 
parameters as discussed in [10].} 
\end{center}
\label{fig:3}       
\end{figure}

Generally, the description of the total charm cross section by MC 
models is not perfect.
For more detailed tests, 
the total inclusive $D^*$ cross
section was  calculated together with the $D^*$ cross section with 
associated dijets \cite{abs074}, as shown in Fig.~5. The CASCADE model
is systematically above the data, while RAPGAP (LO BGF) is below.   

Figure~4 shows the ratio of  $e^-p$ and $e^+p$ $D^{*\pm}$ cross
sections as a function of $\eta(D^{*\pm})$ for $Q^2>1.5$ GeV$^2$.
This ratio exhibits a
trend to increase with increasing $Q^2$ and $x$ \cite{abs563}.
For $Q^2>40$ GeV$^2$, there is $\sim 3\sigma$ difference between
$e^-p$ and $e^+p$ $D^{*\pm}$ cross sections.
According to the Standard Model, charm cross sections
do not depend on the charge of the lepton in $ep$ interactions. 
This difference is assumed to be a statistical fluctuation, and
two sets of the data were combined.  
More $e^-p$ data from HERA II is necessary to
show whether the difference with the $e^+p$ data is indeed a statistical fluctuation.

\section{Extrapolation results}

A popular way to look at double differential charm cross sections 
as functions of $Q^2$ and $x$ is to reconstruct \ff. 
Integrated cross sections in $Q^2$ and $y$ kinematic bins
were extrapolated to the full phase space using HVQDIS based on the Peterson
fragmentation function. 
Usually, several uncertainties
in the extrapolation are considered. The largest uncertainties
are those associated with the AROMA model 
for fragmentation and charm-mass variations ($\pm 0.15$ GeV).

Figure~6 shows the \ff as a function of $x$ at $Q^2$ values between 2 and 500 GeV$^2$.
The data rise with increasing $Q^2$; the rise becoming
steeper at lower $x$.
Comparisons of the ZEUS measurements \cite{abs563} with previous results from
H1  \cite{h1cc} show  good agreement. The data are well described by the
NLO prediction based on the ZEUS NLO QCD fit.

\section{Conclusions}

The production of $D^{*\pm}$ mesons has been measured in DIS at
HERA over a wide kinematic range in $Q^2$, 
from $Q^2=1.5$ GeV$^2$ to $Q^2=1000$  GeV$^2$.
The NLO QCD based on the FFNS gives a good agreement with the data up to 
the highest $Q^2$ range measured. 

At present, no conclusive statement can be made on the applicability of the CCFM
evolution, since other effects related to charm fragmentation and   
the gluon density inside the
proton, are shown to affect the $D^{*\pm}$ cross sections.

More data from HERA II is necessary to increase 
the precision of the measurements 
and to extend the kinematic range of reconstructed charm mesons. 
This should allow to understand
the applicability of the FFNS at high $Q^2$ and the CCFM parton 
evolution at low $x$, as well as  
to understand 
whether there is a difference between $e^-p$ and  $e^+p$ $D^{*\pm}$ data.
 
\begin{figure}
\begin{center}
\resizebox{0.3\textwidth}{!}{%
  \includegraphics{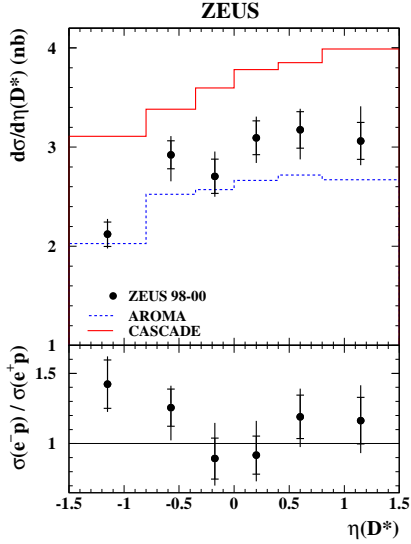} }
\caption{Differential $D^{*\pm}$ cross section as a function of 
$\eta(D^*)$ compared with the AROMA (dashed line)
and CASCADE (solid line) MC programs. Other details as for Fig.1.
}
\end{center}
\label{fig:4}       
\end{figure}

\begin{figure}
\begin{center}
\resizebox{0.3\textwidth}{!}{%
\includegraphics{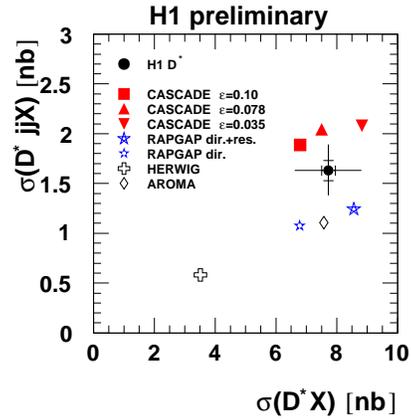} }
\caption{The total inclusive $D^{*\pm}$ cross section 
versus the  $D^{*\pm}$ cross section
with associated dijet compared to MC models. 
The cross section was measured for 
$Q^2>2$~GeV$^2$, $0.05<y<0.7$  
$1.5 < p_T(D^{*\pm}) < 15$~GeV  and $|\eta(D^{*\pm})|<1.5$. For the $D^{*\pm}$ 
cross section
with associated dijets, $E_{T1}> 4$ GeV, $E_{T2}> 3$ GeV and
$-1< \eta (\mathrm{jets}) < 2.5$ were applied.
}
\label{fig:5}       
\end{center}
\end{figure}

\begin{figure}
\begin{center}
\resizebox{0.35\textwidth}{!}{%
  \includegraphics{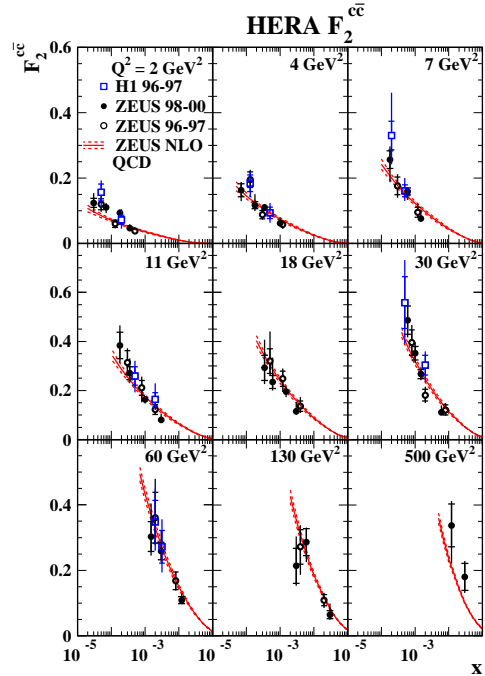}
}
\caption{
The measured  $F_2^{c \bar{c}}$ 
at $Q^2$ values between 2 and
500~GeV$^2$ as a function of $x$. The curves correspond to the
result of the ZEUS NLO QCD fit, where the lower and upper curves show the
uncertainty on the parton distributions for the ZEUS NLO fit. }
\label{f2}       
\end{center}
\end{figure}

%

\end{document}